\magnification\magstep1
\vglue 1cm 
\input mssymb 

\hyphenation{mo-de}

\centerline {\bf AXIOMATIC ANALYTICITY PROPERTIES \rm}
\smallskip

\centerline {\bf AND REPRESENTATIONS  
OF PARTICLES \rm}
\smallskip

\centerline {\bf IN THERMAL QUANTUM FIELD THEORY
\rm}
\smallskip 

\centerline { by }
\smallskip

\centerline { \bf Jacques Bros$^a $ and Detlev Buchholz$^b $ \rm}
\vskip 0.5cm 

\centerline {$^a$ Service de Physique Th\'eorique, CEA-Saclay}
\centerline { F-91191 Gif-sur-Yvette, France}
\smallskip

\centerline {$^b$ II. Institut f\"ur Theoretische Physik,
Universit\"at Hamburg}
\centerline {D-22761 Hamburg, Germany}
\vskip 1cm 

\noindent
ABSTRACT.  We provide an axiomatic framework for Quantum Field
Theory at finite temperature which implies the
existence of general analyticity properties of the $ n $-point
functions; the latter parallel the properties derived from
the usual Wightman axioms in the vacuum representation of
Quantum Field Theory. Complete results are given for the 
propagators, including a generalization of the K\"all\'en-Lehmann
representation. Some known examples of ``hard-thermal-loop
calculations'' and the representation of ``quasiparticles''
are discussed in this general framework. 
\vskip 1cm 


\centerline { \bf 1. INTRODUCTION \rm}
\parskip 5pt 

\noindent In the past fifteen years, there has been an increasing 
interest for
the physics of media at very high temperature; in the latter, 
the basic quantum fields of matter are supposed to manifest
themselves through new properties which highly deserve to be
investigated from a theoretical viewpoint. In particular,
the so-called ``quark-gluon plasma'' has been the object of
a number of theoretical studies and computations [1-3]. These are
generally based on an adaptation of the standard methods of
quantum statistical mechanics to the relativistic system
consisting of the basic fields of quantum chromodynamics.
The fields are supposed to be in an equilibrium state of
infinite volume at temperature $ T=1/\beta $ in a certain
privileged Lorentz frame; the latter fixes the time and space
variables $ (x_0,\vec x) $ and the corresponding Fourier conjugate
variables, namely the energy and momentum $ (\omega,\vec p) $.
\parskip 5pt

The use of the Matsubara \it imaginary-time formalism \rm
(ITF) (see e.g. [4] and references therein)   
has resulted in computations
involving discrete imaginary-energy summations for obtaining
perturbative approximations of the retarded propagators of the
fields in the space of (complex) energy and (real) momentum
variables. In this formalism, the \it quasiparticles \rm are
associated with ``modes'' obeying a (real or complex) 
``dispersion law'' $ \omega = f(\vec p) $, which appear as
poles of the form $ Z(p)/[\omega - f(\vec p)] $ in the retarded
propagators of the fields. Such a structure has been displayed
in the so-called ``hard-thermal-loop calculations'' which aim
to exhibit the very high temperature behaviour of a quark-gluon
plasma; summing the one-loop self-energy contributions 
in a leading approximation at large $ T $ 
yields the following result: the gluon
propagator exhibits two typical poles
$\omega=f_t(\vec p) $ and $ \omega=f_l(\vec p) $, 
interpreted respectively
as ``transverse and longitudinal plasmon modes'' [5,6]. 

The independent elaboration of a formalism which only involves
\it real-time quantities \rm (RTF) has resulted in a double-field
matrix formulation, now called \it thermo-field dynamics \rm
(see e.g. [7] and references therein). This approach introduces as 
preferable basic functions the time-ordered and anti-time-ordered
expectation values.

In the past years, there has been a long debate
about the consistency of the ITF and RTF approaches, 
the latter being both expressed in terms of appropriate versions
of the path-integral formalism; the controversy
has generally been set in this framework, where genuine subtleties  
appear in handling ``time-ordered paths'' in the complex plane
of the time-variable (see e.g. [8]). Although the situation now seems to 
have been clarified 
in favour of the consistency [9], 
one feels a real need for a general structural study of the thermal
Green functions going beyond the time-ordered path technique.   

In the present work, we adopt such a general (non-perturbative) 
viewpoint on Thermal Quantum Field Theory: we suggest that the 
validity and consistency of the ITF and RTF formalisms are
by-products of model-independent structural properties of
the thermal $n$-point functions which follow rigorously from
an appropriate set of general physical principles. As a matter
of fact, we will show that one can define an axiomatic program
for studying the analytic and algebraic properties of the $ n $-point
functions of fields in a thermal equilibrium state. This program
is completely similar to the one which has been developped in the
sixties for Quantum Field Theory in the vacuum state within the
Wightman axiomatic framework and which has led to such important
structural insights as the PCT and Spin-Statistics Theorems,
collision theory, dispersion relations and the 
Osterwalder-Schrader Theorem (see [10-13] and references therein). 

The main 
discrepancy of our axiomatic starting point with respect to the 
familiar Wightman axioms will consist in releasing 
the Lorentz covariance properties and replacing the \it
spectral condition \rm by an appropriate formulation
of the {\it KMS-condition\/}\rm: this idea relies on the basic
analysis of [14], completed by the results of [15,16] 
which display the KMS-property as being a general 
criterion for systems in equilibrium states. Moreover, in spite of 
the breaking of Lorentz covariance due to the thermal bath, the
basic role played by the causality cone $ V^+ = \{ x=(x_0,\vec x); 
x_0 > | \vec x | \} $ for quantum field systems implies that a 
(stronger) \it relativistic form of the KMS-condition \rm can be
justified [17], as a remnant of the \it relativistic spectral
condition. \rm

In this preliminary work, we shall present substantial results
for the structure of the two-point functions; only partial
results will be given as far as the general
program of $ n $-point functions is concerned. However, the case
of the thermal two-point functions being already of physical 
interest, we give a special importance to a K\"all\'en-Lehmann-type
representation (already presented in [18,19])  
which we are able to derive from the general
principles; among other advantages, this representation opens
a new possibility for the characterization of particles in
Thermal Quantum Field Theory.

After having presented our axiomatic framework for the 
representations of Quantum Field Theory in thermal
equilibrium states in Sec.2, we devote Sec.3 and 4 
to the pure implications of locality: this part of the program
reproduces in the setting of Thermal Quantum Field Theory 
properties of 
general field theory in the vacuum 
state. It concerns:                                           

\noindent a) in Sec.3, the basic analyticity properties of $n$-point Green
functions in the space of complex energy and momentum variables, 

\noindent b) in Sec.4, the derivation of K\"all\'en-Lehmann-type integral
representations for the two-point commutator and retarded functions
which are valid \it in the absence of Lorentz covariance and of
spectral condition. \rm

The implications of the KMS-condition will be studied in Sec.5. In
particular, we shall exhibit the resulting \it double analytic structure
\rm of the two-point functions in the time and energy variables
and the relations between the Fourier transforms of the retarded and
time-ordered functions which replace the usual ``coincidence
relations'' implied by the spectral condition. 
The derivation of similar properties for the $ n $-point functions
will only be initiated there and mentioned 
as a further important part of our program to be
implemented. 

In Sec.6, the following complements on the structure of two-point
functions
will be presented:

\noindent a) incorporating the KMS-condition into 
the K\"all\'en-Lehmann-type formula
of Sec.4 in order to provide a corresponding representation for 
the two-point correlation function itself,

\noindent b) completing the results of Sec.5 by analyticity properties in the
\it complex space and time variables \rm which result from our
relativistic form of the KMS-condition, 

\noindent c) studying the basic Feynman-type operations on 
two-point functions and
illustrating all the previous results on the perturbative examples
mentioned at the beginning of this introduction, 

\noindent d) giving a comparative discussion of two different characterizations
of the notion of particle in Thermal Quantum Field Theory.
\vskip 1cm

\centerline { \bf 2. THE AXIOMATIC FRAMEWORK \rm} 
\parskip 5pt

\noindent The adaptation of the Wightman axiomatic framework to the case
of quantum fields in a thermal equilibrium state $ \Omega_\beta $
of temperature $  T= 1/ \beta $ relies 
on the following well-known ideas.  For
simplicity, we consider the case of a single hermitian field
$ \phi(x) $, namely a system of ``observables''
$ \phi(f) = \int \phi(x) f(x) dx $ 
depending continuously on test-functions $ f $ on Minkowski space
(taken for convenience in the Schwartz space $ {\cal S}({\Bbb    R}^4) $).
These observables form an algebra which satisfies the general
axiom of ``locality'' or ``local commutativity'', namely:
$ [\phi (f), \phi (g)] = 0 $  for all pairs of test-functions $ (f,g) $ 
whose supports are mutually space-like separated in $ {\Bbb    R}^4 $. 

This axiom, which expresses the principle of Einstein causality, 
is independent of the \it representation \rm in which a 
system is described; it therefore holds in the Hilbert space
$ {\cal H}_{\beta} $ 
of thermal states in which the field
observables $ \phi (f) $ are supposed to act, 
as operators defined on a suitable dense domain generated by
the state vector $ \Omega_\beta $.
It is understood that $ \Omega_\beta $ 
determines a fixed Lorentz frame, i.e.\ a distinguished set of
time and space variables; the corresponding unitary
representation $ U_t $ of the time-translation group in $ {\cal
H}_\beta$
(i.e.\ the evolution operator group) leaves $ \Omega_\beta $ invariant,
since an equilibrium state is stationary. We shall only consider
here the case when  $ \Omega_\beta $ is also invariant under the  
(unitary) representation of space-translations $ U_{\vec a} $ in
$ {\cal H}_\beta $. The action of the space and time-translations on the
field is, as usual: $ \phi(x_0+t,\vec x+\vec a)= U_t U_{\vec a}
\phi(x_0,\vec x) U_{\vec a}^{-1} U_t^{-1}  $.
The previous axioms do not differ from the corresponding Wightman
axioms in the Hilbert space $ {\cal H}_{vac} $ generated by the vacuum
state $ \Omega_{vac} $; however, there are no unitary operators
which implement Lorentz transformations in $ {\cal H}_\beta $.
\parskip 5pt

In order to complete our axiomatic framework
in $ {\cal H}_\beta $, we shall introduce the ``correlation functions''
or ``Wightman functions'' $  {\cal W}^{(\beta)}_n(x_1,\ldots,x_n) = 
<\Omega_\beta,\phi(x_1)\cdots\phi(x_n)\Omega_\beta>  $
and adopt the viewpoint of
the reconstruction theorem [10].  A Thermal Quantum Field Theory
is thus entirely specified by the knowledge 
of the corresponding set of tempered
distributions $ \{{\cal W}_n^{(\beta)} ;n\in {\Bbb    N}\} $
which have to satisfy the usual \it positivity conditions \rm 
of Wightman functions [10]. (It can equivalently be presented
as \it the representation associated with a certain 
positive functional \rm on the Borchers-Uhlmann algebra [20,21]
of terminating sequences of test-functions 
$ \{f_0,f_1(x_1),\ldots,f_n(x_1,\ldots,x_n),0,\ldots\}$.)

Apart from the properties which express the previous axioms, 
namely:

\noindent \bf i) Locality: \rm
$$\;\;\;\;\;\;
 {\cal W}^{(\beta)}_n(x_1,\ldots,x_i,x_{i+1},\ldots, x_n) = 
 {\cal W}^{(\beta)}_n(x_1,\ldots, x_{i+1}, x_i\ldots, x_n) 
\eqno{(1)}$$
for $ x_i-x_{i+1} \not\in \overline {V^+} \cup \overline {V^-} $ 
with  ${V^-}={-V^+}, $

\noindent \bf ii) Translation Invariance: \rm 
$$\;\;\;\;\;\;
 {\cal W}^{(\beta)}_n(x_1,x_2, \ldots, x_n) = 
{\cal W}^{(\beta)}_n(x_1 + a,x_2 + a,\ldots,  x_n + a) \eqno{(2)} $$
for all $a$ in ${\Bbb R}^4$, 
the thermal Wightman functions should satisfy the following 
analyticity properties which replace those implied by the spectral
condition in the vacuum case. These properties are the mathematical
expression of the fact that the state $ \Omega_\beta $  is in thermal 
equilibrium [14-16].

\noindent \bf iii) KMS-condition: \rm 

\noindent Let $ e_0 $ be the unit vector of the time-axis. For every
$  n \in {\Bbb  N} $ and every pair $ (I,J) $ of \it ordered
sets \rm  $ I= \{1,\ldots,m\}, J=\{m+1,\ldots,n\} $, 
there exists a tempered distribution
$ F_{IJ}( x_1 ,\ldots,x_n ;t) $ 
which is holomorphic with respect to the complex variable $ t $ 
in the strip $ - \beta < Im t < 0  $  and admits distribution 
boundary values on $ \{t \in R \} $ and  $ \{t \in R - i \beta \} $
(still denoted by $ F_{IJ} $) satisfying the following conditions:
$$ F_{IJ}(x_1,\ldots,x_n;t) ={\cal W}^{(\beta)}_n(\{x_i+te_0\}_{i\in I},
\{x_j\}_{j\in J}), \eqno{(3)}$$
$$ F_{IJ}(x_1,\ldots,x_n;t-i\beta) ={\cal W}^{(\beta)}_n(\{x_j\}_{j\in J}
,\{x_i+te_0 \}_{i\in I}).  \eqno{(4)}$$

Let us introduce the Fourier transforms of the Wightman
functions, which (in view of ii)) can be written as
$ \hat {\cal W}^{(\beta)}_n(p_1,\ldots,p_n)\ \delta(p_1+\cdots+p_n) $,
with $ p_l=(\omega_l,\vec p_l),1\le l\le n $. Then 
the distributions $ \hat {\cal W}_n^{(\beta)} $ on the linear manifold 
$ p_1 +\cdots + p_n = 0 $ satisfy the following property which
is equivalent to iii):
\parskip 5pt

\noindent \bf iii') KMS-condition in the energy variable: \rm

\noindent For each pair $ (I,J) $, the following identity holds:
$$ \hat {\cal W}^{(\beta)}_n(J,I) = 
 e^{-\beta \omega_I}\hat{\cal W}^{(\beta)}_n(I,J), \eqno{(5)}$$

\noindent
where $ \omega_I = \sum_{i \in I} \omega_i  $
and $ \hat{\cal W}^{(\beta)}_n(I,J) $  stands for 
$ \hat{\cal W}^{(\beta)}_n(\{p_i\}_{i\in I}, \{p_j\}_{j\in J}) $.
\smallskip

\noindent 
\it Remark:  \rm  If we introduce the commutator functions
$$ \hat {\cal C}^{(\beta)}_n(I,J)=
\hat {\cal W}^{(\beta)}_n(I,J) -
  \hat{\cal W}^{(\beta)}_n(J,I) \eqno{(6)} $$

\noindent 
it follows from Eq.(5) that $\hat {\cal W}^{(\beta)}_n$ 
is ``essentially'' determined   
from $\hat {\cal C}^{(\beta)}_n$ 
(i.e. determined up to a part which
factorizes $ \delta (\omega_I) $) by the formula:
$$ (1-e^{-\beta \omega_I})\hat {\cal W}^{(\beta)}_n(I,J)=
\hat {\cal C}^{(\beta)}_n(I,J). \eqno{(7)}$$

\noindent
We notice that this formula replaces the usual one expressing
the spectral condition in the case of the vacuum state:  
$$\hat {\cal W}^{(vac)}_n(I,J)=
\theta(\omega_I)\hat {\cal C}^{(vac)}_n(I,J) \eqno{(8)}$$

\noindent
($ \theta $   denoting the Heaviside step-function).
\smallskip

\noindent
In the limit of zero temperature ($ \beta \to \infty $),
Eq.(7) tends to Eq.(8);  the ``inverse Bose-Einstein factor''
 $(1-e^{-\beta \omega_I})$ 
in Eq.(7) implies (since $\hat {\cal W}^{(\beta)}_n$ is a tempered 
distribution) that 
$\hat {\cal W}^{(\beta)}_n(I,J)$ has an \it exponentially decreasing
behaviour \rm of the form $ e^{-\beta| \omega_I |} $
for $ \omega_I $ tending to
$ - \infty $, instead of being equal to zero for $ \omega_I < 0 $
as $\hat {\cal W}^{(vac)}_n(I,J)$.  The occurence 
of negative energy contributions in
$\hat {\cal W}^{(\beta)}_n$ corresponds to the possibility of 
``extracting energy from the ambient
thermal bath''. The exponential suppression of these 
contributions is a remnant of the \it condition of positivity of 
energy \rm in the vacuum case.
In fact, in the relativistic theory the latter is true in all Lorentz 
frames (\it relativistic spectrum condition\rm); this    
implies a more stringent support condition
for $\hat {\cal W}^{(vac)}_n(I,J)$  than the condition  $ \omega_I > 0  $
exhibited by Eq.(8), namely: 
$$ {\rm supp}\ \hat {\cal W}^{(vac)}_n(I,J) \subset
\{ (p_1,\ldots,p_n); p_I = -p_{J} \in
\overline{V^+} \},\ \eqno{(9)}$$
where $ p_I = \sum_{i\in I} p_i $ and $ p_J $ is defined analogously. 

A corresponding relativistic form of the 
KMS-condition has been justified in the 
analysis of [17] based on the general principles of ``Local Quantum
Physics'' [22];  this analysis gives a firm background to the 
following properties (stronger than iii), iii')) of the Wightman
functions $ {\cal W}_n^{(\beta)}(I,J) $. 
\smallskip

\noindent \bf iv) Relativistic KMS-condition:\rm 
\smallskip

\noindent For each pair $ (I,J) $, the distribution $ F_{IJ} $ 
of condition iii) admits
an analytic continuation (which we still call) 
$ F_{IJ}(x_1,\ldots,x_n;z) $ with respect to the
complex four-vector variable $ z = (z_0, \vec z )  $ in the
following tube-domain:
$ T_\beta = \{ z \in {\Bbb C}^{\, 4}; {\rm Im} z \in V^-, 
{\rm Im} z + \beta e_0 \in V^+\}. $

\noindent
The boundary value equations (3) and (4) can then be replaced
respectively by:
$$ \lim _{\epsilon \to 0, \epsilon \in V^+} F_{IJ}(x_1,\ldots,x_n;x -i\epsilon) =  {\cal W}^{(\beta)}_n(\{x_i+x \}_{i\in I}; \{x_j\}_{j\in J}), \eqno{(10)}$$
$$ \lim _{\epsilon \to 0, \epsilon \in V^+}
F_{IJ}(x_1,\ldots,x_n;x-i\beta+i\epsilon) =  {\cal
W}^{(\beta)}_n(\{x_j\}_{j\in J};\{x_i+x \}_{i\in I}). 
\eqno{(11)}$$

\noindent
In the space of energy and momentum variables, condition iv) can
be equivalently expressed by iii')
supplemented by the following condition iv'):
\smallskip

\noindent \bf iv') Essential support conditions in p-space: \rm 
\smallskip

\noindent 
For each pair $ (I,J) $, the distribution $ \hat {\cal W}^{(\beta)}_n(I,J) $ 
admits the cone
$ \{(p_1,\ldots,p_n);p_I=-p_J\in \overline{ V^+} \} $ as a majorant of
its ``essential support in the sense
of exponential decrease''. A more precise formulation of this
condition is that the products of 
the tempered distribution $ \hat {\cal W}^{(\beta)}_n(I,J) $
by the (smooth) functions  
$ e^{{\beta\over2} \big( (1+{\vec p}_I^2)^{1\over2} -
\omega_I \big) } $ and $ e^{-\beta \omega_I} $ are  
required to be also tempered.
 
For completeness, we should also mention an additional postulate
whose role is to restrict our framework to the case of
equilibrium states $ \Omega_\beta $ which are \it pure phases. \rm 

\noindent \bf v) Time-clustering postulate: \rm

\noindent 
For each pair $ (I,J) $, one has in the limit $ \vert t\vert \to \infty
$ 
(with the notation used in Eq.(3)): 
$$ F_{IJ}(x_1,\ldots,x_n;t) = {\cal W}_m^{(\beta)}
(\{x_i\}_{i\in I})\  {\cal W}_{n-m}^{(\beta)}
(\{x_j\}_{j\in J}) + o(1).  \eqno (12) $$
\vskip 1cm

\centerline { \bf 3. ANALYTICITY PROPERTIES}  
\centerline { \bf IN THE ENERGY AND MOMENTUM VARIABLES} 
\smallskip

\noindent Recently, a procedure for studying the usual 
$ n $-point retarded and advanced functions 
and possibly introducing generalized retarded functions 
characterized by various analyticity properties in the 
energy variables has been presented 
in the traditional imaginary-time formalism of
Thermal Quantum Field Theory [23]. 
Here, we would like to emphasize that a general understanding of 
the algebraic and analytic properties of these Green functions 
follows from the Wightman axiomatic approach of quantum field 
theory, and that the latter allows one to control exactly
which results are identical to those of the vacuum case
and which ones are different.

We devote this section to the \sl introduction of $ n $-point 
thermal Green functions enjoying analyticity properties
in the energy and momentum variables. \rm Their definitions
and algebraic study in the previous axiomatic framework 
and the (correlated) derivation of their
analyticity properties as 
\sl pure consequences of the axiom of locality \rm 
are completely identical to those which 
have been given for the 
vacuum representation of Quantum Field Theory.
This structure has been the object of a number of works 
in the past [24-28]. 

The basic idea is that for each $ n $, 
one can define certain privileged combinations $ r_\alpha $ of permuted 
n-point Wightman functions multiplied by appropriate
products of step-functions of the (differences of) time-variables.
Each distribution $ r_\alpha $ enjoys remarkable 
\it support properties \rm 
which are \it implied by locality, \rm namely the support of $ r_\alpha
$ is contained in a 
\it (salient) convex Lorentz invariant cone \rm
$ \Gamma_\alpha $ of the space of vector differences $ x_i-x_j $. 
Then, in view of a basic result of complex 
analysis (see e.g. [10,29]), the Fourier transform 
of $ r_\alpha $ is the boundary value
on the reals of a holomorphic function $ \tilde r_\alpha $
in a tube-domain $ T_\alpha $  
of the \it complex energy-momentum space \rm which is of the 
following form:  $ \{k=(k_1,\ldots,k_n);
k_1+\cdots +k_n=0; Im k \in \tilde \Gamma_\alpha \} $,
where $ \tilde \Gamma_\alpha $ (called the 
\it basis of the tube\  \rm  $ T_\alpha $) 
is the dual cone of $ \Gamma_\alpha $
(i.e. the set of vectors $ q $ such that $ qx= q_1x_1+\cdots+ q_nx_n $
is positive for all $ x $ in $ \Gamma_\alpha $). 
Each of these distributions $ r_\alpha $ (or holomorphic
functions $ \tilde r_\alpha $)
will be called a \it generalized retarded function. \rm 
It is the expectation value $ <\Omega_\beta,R_\alpha \Omega_\beta> $
of a corresponding \it generalized retarded operator \rm 
$ R_\alpha(x_1,\ldots,x_n) $. 
The generalized retarded operators (resp. functions) satisfy two sets of basic 
algebraic relations, namely:
\smallskip

\noindent a) discontinuity relations between the various $ R_\alpha $ (resp.
$ r_\alpha $); in this connection,   
\ \it generalized absorptive parts \rm are introduced; 
\smallskip

\noindent 
b) relations between each $ R_\alpha $ and the \it (anti) time-ordered
operator products. \rm 
\smallskip 

\noindent
\bf The two-point function:\rm 
\smallskip

\noindent Let $ {\cal C} (x) $   be the commutator function, i.e.,  
$ {\cal C}(x_1-x_2) = {\cal W}_2^{(\beta)} (x_1,x_2) -
{\cal W}_2^{(\beta)} (x_2,x_1) $. 
Locality implies that  $ \hbox{supp } {\cal C} \subset \overline {V^+} 
\cup \overline {V^-}. $ 
There are two distributions $ r_\alpha $ in this case, namely
the ``retarded and advanced functions'', defined formally as
$ r(x)= i\theta(x_0) {\cal C}(x) $ and $ a(x)= -i\theta(-x_0) {\cal C}(x)
$.
\parskip 5pt

The relation $ r-a= i\cal C $ corresponds to the splitting of the
support of $ \cal C $ into its two convex components, since
$ \hbox{supp } r \subset \overline {V^+} $ and 
$ \hbox{supp } a \subset \overline {V^-}. $
(Note that this splitting is defined up to a distribution with support at the
origin).

This splitting is in turn equivalent to the following one for the   
Fourier-transformed quantities: 
$ i \tilde {\cal C}(p)= \tilde r(p) -\tilde a(p) $, 
where $ \tilde {\cal C} $ is usually called \it
spectral function \rm [1-8] and $ \tilde r $ and $ \tilde a $
are the boundary values of holomorphic functions (denoted similarly
by) $ \tilde r (k) $ and $ \tilde a(k) $ $ (k=p+iq) $ in the respective
tube-domains $ T^+ = \{ k=p+iq ; q \in V^+ \} $ and $ T^- = -T^+ $
of $ {\Bbb C}^{\, 4} $.  
Conversely, given a  
holomorphic function in $ T^+ \cup T^- $ (with at most a tempered
behaviour at infinity and near the reals), the spectral function
obtained by taking the difference of the corresponding boundary values
defines a commutator function which does satisfy locality. According
to the common use, the holomorphic function $ \tilde r(k) $
(or $\tilde a(k) $) is then called the \it propagator \rm whose
associated spectral function is $ \tilde {\cal C}(p) $. 
(In view of the Hermitian character of the field, 
$ \tilde {\cal C}(p) $ is also the imaginary part of 
$ \tilde r(p) $ and is commonly described as such, although
its characterization as \it the discontinuity of the holomorphic 
function $\tilde r(k) $ across the reals \rm is more substantial). 

Let $ \tau (x) $, $ x=x_1-x_2  
$ (resp.\ $ \overline \tau (x) $) be the two-point
time-ordered (resp.\ anti-time-ordered) function. The following relations
hold: 

 $$ r=i(\tau -{\cal W}')=-i(\overline \tau -{\cal W}),\ \
 \ \  a=i(\tau -{\cal W}) =-i(\overline \tau -{\cal W}'), \eqno{(13)} $$  
 where  $ {\cal W}(x_1-x_2)={\cal
W}_2^{(\beta)}(x_1,x_2)  $ 
and $ {\cal W}'(x_1-x_2)=
{\cal W}_2^{(\beta)}(x_2,x_1). $

\noindent
\bf  Examples: \rm 

\noindent
1) \sl The free field propagators: \rm

\noindent
For any free scalar field of mass $ m $, the commutator function
$ {\cal C}^{(m)} $
and therefore the associated 
retarded and advanced functions $ r^{(m)} $ and $ a^{(m)}  $ are 
independent of the representation generated by the thermal state 
$ \Omega_\beta $. In fact, in this case $ {\cal C}^{(m)}(x)  $ is a structural
function of
the field algebra determined by the c-number
commutation relations of the field. Therefore, for every thermal
representation with temperature $ \beta^{-1} $, one has
(as in the vacuum representation):
$\  {\cal C}^{(m)}(x) = {1\over (2\pi)^2}\int_{{\Bbb  R}^4} e^{-i px} 
\ \tilde {\cal C}^{(m)} (p)\ dp
$, with
$$ \tilde{\cal C}^{(m)}(p) = 
{1\over 2\pi}\epsilon(\omega) \delta(\omega^2 - {\vec p}^{\, 2} - m^2)
\eqno{(14)}$$
and 
$$ \tilde r^{(m)}(k)\  (\hbox{resp.}\ \ \tilde a^{(m)}(k) ) = 
-{1\over 4\pi^2}{(k_0^2- \vec k^2 -m^2)}^{-1}   \eqno(15) $$ 
for $k = (k_0,\vec k) = p+iq \in T^+  $ (resp.\ ) $  k \in T^-) $.

\noindent
2) \sl The Weldon-Pisarski (WP) propagators: \rm

\noindent
The following expressions of the
high-temperature transverse and longitudinal
gluon propagators have been obtained by the resummation of
dominant contributions at small $ \beta $ from the one-loop
self-energy diagrams: 
{$$ \Delta_t(k) = \left[k_0^2 - \vec k^2 - {M^2 \over 2 \vec k^2 } 
\left(k^2_0 + {k_0(\vec k^2 - k_0^2) \over 2 \sqrt{\vec k^2 }}
\log\left({k_0 + \sqrt{\vec k^2 } \over k_0 - \sqrt{\vec k^2}}\right)\right)
\right]^{-1}, \eqno{(16)} $$ 

$$ \Delta_l (k) = { \vec k^2 \over {k_0^2 -\vec k^2}} 
\left[  \vec k^2 + M^2 \left(1 - {k_0 \over 2 \sqrt{\vec k^2}}
\log\left({k_0 + \sqrt{\vec k^2} \over k_0 - \sqrt{\vec k^2}}\right)\right)
\right]^{-1}. \eqno{(17)} $$
\noindent
In these expressions, $ M $ denotes the ``Debye screening mass''
which is directly related to the plasma frequency.
The propagators $ \Delta_t $ and $ \Delta_l $ 
can be checked to be holomorphic in the tubes
$ T^+ $ and $ T^- $  in the privileged sheet of the logarithm
which defines the \it physical sheet \rm of these functions.
Therefore the associated two-point functions satisfy
locality (in this connection, see [30] for a detailed numerical
analysis of these examples and further comments 
on the construction of causal propagators).
\smallskip

\noindent
\bf The $ n $-point functions: \rm
\smallskip

\noindent For each $ n $, the distributions $ r_\alpha $ are labelled by a 
``cell-function'' $ \alpha $ which represents a geometrical cell
$ \gamma_\alpha $ (namely a polyhedral cone)
in the space $ H^{(n)} = \{h=(h_1,\ldots,h_n)\in {\Bbb R}^n;
h_1+\cdots+h_n=0\} $
deprived from all the hyperplanes $ H^{(n)}_{I,J} $ with equation
$ h_I=-h_J=0 $. Here, $ (I,J) $ denotes an \it arbitrary \rm partition of 
$ \{1,\ldots,n\} $ and we use the notation $ h_I= \sum_{i\in I}h_i. $ 
A cell $ \alpha $ is defined as the ``sign-valued function''
$ I \to \alpha(I)= +\  \hbox{or}\  - $ such that $ \gamma_\alpha=
\{h \in H^{(n)}; \forall I\subset\{1,2,\ldots,n\}, 
\alpha(I)h_I>0\} $. 
Two cells $ \alpha_1 $ and $ \alpha_2 $ are called ``adjacent
cells along the face $ H^{(n)}_{I_0,J_0} $'' if they only differ on 
the complementary sets $ I_0 $ and $ J_0 $, namely if $ \alpha_1(I_0 ) =  
-\alpha_1(J_0) = -\alpha_2(I_0) = \alpha_2(J_0) $. 
The corresponding generalized retarded operators or functions
will also be said to be adjacent.
\smallskip

Each generalized retarded function $ \tilde r_\alpha $ is holomorphic
in the tube $ T_\alpha $ whose conical basis $ \tilde \Gamma_\alpha $
in $ \{q= Im k =(q_1,\ldots,q_n); q_1+\cdots+q_n = 0\} $ is defined
by the set of conditions $ \alpha(I) q_I \in V^+ $ (for all
proper subsets $ I $ of $ \{1,2,\ldots,n\} $) [27,28].
Each ordinary retarded function $ r^{(i)} $ corresponds to
the cell $ \alpha $ such that $ \alpha(\{j\})=+ $ for all 
$ j \in \{1,2,\ldots,n\} $ except for $ j=i $. 
\smallskip

\noindent
\sl Discontinuity relations:\rm 
\smallskip

\noindent
They are generated by the following
set of relations which connect all the pairs of adjacent
generalized retarded operators [25,26]: 
for every partition $ (I,J) $ of $ \{1,\ldots,n\} $
and for every pair of cells  $ ( \alpha_1,\alpha_2) $ which are
adjacent along the face $ H^{(n)}_{I,J} $, there holds:
$$ R_{\alpha_1} - R_{\alpha_2} = i[R_{\alpha^{(I)}} , R_{\alpha^{(J)}}], 
\eqno{(18)}$$

\noindent
where $ \alpha^{(I)},\alpha^{(J)} $ denote the common restrictions  of
the cell-functions $ \alpha_1, \alpha_2 $ respectively 
to the proper subsets of $ I $ and of  $ J $, and $ R_{\alpha^{(I)}},
R_{\alpha^{(J)}} $ are the corresponding generalized retarded operators
in the space of the variables labelled by the elements of $ I $ 
and $ J $. Correspondingly, there holds the following set of relations,
which generalize the relation $ r-a= i{\cal C} $ of the two-point
function: 
$$ r_{\alpha_1} - r_{\alpha_2} = i\left(\langle \Omega_\beta , 
R_{\alpha^{(I)}}R_{\alpha^{(J)}}\Omega_\beta \rangle  
- \langle \Omega_\beta , R_{\alpha^{(J)}}R_{\alpha^{(I)}}\Omega_
\beta \rangle \right).   \eqno{(19)} $$

\noindent
In the latter, the terms at the r.h.s. are interpreted as
\it generalized absorptive parts \rm in the respective
channels $ (I,J) $ and $ (J,I) $.
\smallskip

\noindent
\sl Steinmann relations [24]: \rm 
\smallskip

\noindent
The generalized retarded operators $ R_{\alpha} $ are \it not 
linearly independent. In fact, 
if two adjacent pairs $ (\alpha_1,\alpha_2) $ and
$ (\alpha_1',\alpha_2') $ along the same face $ H^{(n)}_{I,J} $
admit the same restrictions $ \alpha^{(I)},\alpha^{(J)} $,
the following corresponding relation 
holds:
$ R_{\alpha_1} - R_{\alpha_2} = R_{\alpha_1'} - R_{\alpha_2'} $. 
\smallskip

\noindent
\sl Relations with the time-ordered operator products: \rm
\smallskip 

\noindent
Let $ T(I) $ denote the time-ordered product of the 
$ n_I $ field-operators $ \phi(x_i) $
for $ i \in I \subset \{1,2,\ldots,n\} $. 
For each cell $ \alpha $
the following expression of $ R_\alpha $ is valid [28]:
$$ (-i)^{n-1}\ R_\alpha(x_1,\ldots,x_n) = 
T(\{1,2,\ldots,n\}) + {\sum}^{(\alpha)}
(-1)^{r-1}\ T(I_1)\ldots T(I_r), \eqno{(20)} $$
\noindent
where the sum $ \sum^{(\alpha)} $ runs over all \it ordered \rm
partitions $ (I_1,\ldots,I_r), 2\le r\le n, $ 
of $ \{1,2,\ldots,n\} $ such that $ \alpha(I_1) =
\alpha(I_1 \cup I_2) = \alpha(I_1 \cup\ldots\cup I_{r-1}) = - $ . 
Similar relations involving the 
anti-time-ordered operator products
can also be written. Eq.(20) generalizes the relations (13)
of the two-point function.
\vskip 0.7cm

\centerline {\bf 4. INTEGRAL REPRESENTATION } 
\centerline {\bf OF THE TWO-POINT SPECTRAL FUNCTION }
\parskip 5pt

\noindent We present here an integral representation of 
the thermal two-point
commutator and spectral functions 
$ {\cal C}(x) $ and $ \tilde {\cal C}(p) $ 
which is \it  a pure consequence of locality; \rm 
from a technical viewpoint, it also relies on the fact
that $ \tilde {\cal C}(p)/\omega $ has to be a (positive and even) 
\it measure \rm as a by-product of the 
positivity and KMS conditions (see Sec.5). 
Our representation is 
characterized by a certain \it ``weight-function'' \rm
which plays the same role as that of the K\"all\'en-Lehmann
representation of the vacuum two-point function.
This weight-function can also be reconstructed from
$ {\cal C}(x) $ by a simple inversion formula.
\smallskip  
\noindent
\bf Proposition: 
\smallskip  
\noindent
{\it i) The following integral representation holds:
$$ {\cal C}(x)= \int_0^{\infty} dm\  D(\vec x;m)
\ {\cal C}^{(m)}(x); \eqno{(21)} $$
in the latter, the weight-function $ D(\vec x;m) $ is a tempered
distribution with support in 
$ {\Bbb   R}^3 \times \overline{{\Bbb   R}^+} $ 
which is uniquely defined and computable in terms of $ {\cal C} $
by the following inversion formula
$$ D(\vec x;m)= 2i\pi {\partial\over\partial m} \big( \theta (m) \int_{-\infty}
^{+\infty} dx_0  x_0 J_0(m \sqrt {x^2})\  {\cal C}(x) \big) , \eqno{(22)} $$
where $J_0 $ is the zeroth-order Bessel function of the
first kind. 

\noindent
ii) The following corresponding representation applies to the spectral 
function:
$$ \tilde {\cal C}(\omega,\vec{p} \, ) = 
{1\over(2\pi)^{3\over2}} \epsilon(\omega)
\int_{{\Bbb   R}^3} d\vec u \int_0^\infty ds\ \delta(\omega^2 -
{(\vec p-\vec u)}^2 -s)\  \tilde \rho(\vec u,s), \eqno{(23)} $$
where
$$ \tilde \rho(\vec u,s)={1\over2\ (2\pi)^{5\over2} 
\sqrt s} \int_{{\Bbb   R}^3} 
d\vec x\  e^{-i\vec u \vec x} D(\vec x;\sqrt s) 
\eqno{(24)} $$
belongs to $ {\cal S}'({\Bbb   R}^4) $, with
$ \hbox{ supp}\  \tilde\rho \subset {\Bbb   R}^3 \times 
\overline{{\Bbb  R}^+} . $ }  \rm

Similar results were first stated by Gervais and Yndurain
in some unpublished paper [31]; Eq.(23) can be interpreted as
a Jost-Lehmann-Dyson representation (see [12]
and references therein) in the absence of spectral 
condition. The proof which we present here is
self-contained; the missing technical details of distribution 
theory will be given elsewhere [32].
\smallskip 

\noindent
a) \sl Gauss-type transforms in the time and energy-variables: \rm 
we first associate with the (4d-) Fourier pair $ ({\cal C},
\tilde {\cal C}) $ the following (3d-) Fourier pair
$ (\Psi(\vec x;\lambda),\tilde {\Psi}(\vec p;\lambda)) $ 
$$ \Psi(\vec x;\lambda) = i \int dx_0\ e^{-x_0^2\over 4
\lambda} x_0\  {\cal C}(x_0, \vec{x} ),  \eqno (25) $$
$$ \tilde {\Psi}(\vec p;\lambda) = (2\lambda)^{3\over2}
\int d\omega\  
e^{-\lambda\omega^2} \omega\  \tilde {\cal C}(\omega,\vec{p} ).  
\eqno (26) $$

\noindent
It can be checked that $\Psi $ and $\tilde \Psi $ are tempered
distributions (resp. in $\vec x $ and $\vec p $) which are
(slowly-increasing) holomorphic functions of $\lambda $ in the
complex half-plane $ {\Bbb  C}^+ = \{ \lambda; Re \lambda >0\} $.
Moreover, the mapping $\tilde {\cal C} \to \tilde {\Psi} $ and
(thereby) $ {\cal C} \to {\Psi} $ are one-to-one since the integral
transformation (26) can be considered as a Laplace-transformation
in the variable $ \omega^2 $ and therefore inverted (here, one makes
use of the fact that $ \tilde {\cal C}(\omega,\vec p)\over \omega $ 
is an even measure in $ \omega $). 
\smallskip

\noindent
b) \sl The transforms $ \Phi $ and $\tilde {\Phi} $: \rm  We define
$$ \Phi (\vec x;\lambda) = e^{\vec x^2\over 4\lambda}\ \Psi
(\vec x;\lambda).\ \eqno (27)  $$
\it Locality implies \rm that the distribution-valued holomorphic
function $ \Phi (\vec x,\lambda) $ is 
\it tempered \rm like $ \Psi (\vec x,\lambda) $.
By inverting Eq.(27), we obtain the following relation between the
Fourier transforms $\tilde {\Phi}(\vec p; \lambda) $ and 
$ \tilde {\Psi}(\vec p;\lambda) $ of $\Phi $ and $\Psi $: 
$$ \tilde {\Psi}(\vec p;\lambda)= \left( 
{\lambda\over\pi}\right)^{3\over2} 
\int_{{\Bbb R}^3} d\vec{u} \, e^{-\lambda(\vec p-\vec u)^2}
\tilde {\Phi}(\vec u;\lambda),\  \eqno (28)  $$
the latter being valid for all $ \lambda $ in $ {\Bbb   C}^+$.
\smallskip

\noindent
c) \sl The ``weight-functions'' $ \rho $ and $ \tilde {\rho} $: \rm 
we call respectively $ {\rho}(\vec x;s) $ and $ \tilde {\rho}(\vec p;s)
$ the inverse Laplace-transforms of $\Phi $ and $\tilde {\Phi} $ 
with respect to the variable $\lambda $; $ ({\rho},\tilde {\rho}) $
is a (3d-) Fourier pair of tempered distributions on $ {\Bbb   R}^4 $
with support contained in $ {\Bbb   R}^3 \times \overline
{{\Bbb R}^+} $. For our needs, we write explicitly  the
mappings $ {\Phi} \to {\rho} $ and $ \tilde {\rho} \to \tilde {\Phi} $:
$$ {\rho}(\vec x;s) = {1\over{2\pi}} \int_{-\infty}^{+\infty} 
d\nu\ e^{i\nu s} {\Phi} 
(\vec x;i\nu),\ \eqno (29)  $$
$$ \tilde {\Phi}(\vec p;\lambda)= \int_0^{+\infty} ds\ e^{-\lambda s}
\tilde {\rho}(\vec p;s).\ \eqno (30)  $$ 
\smallskip

\noindent
d) \sl The integral representations (21), (23): \rm 
by plugging Eq.(30) into Eq.(28), we obtain
$$ \tilde{\Psi}(\vec p;\lambda)= \left( 
{\lambda\over\pi}\right)^{3\over2} 
\int_{{\Bbb  R}^3} d\vec{u} \, \int_0^\infty ds\ 
e^{-\lambda \big( ( \vec p-\vec u)^2 + s \big) } {\rho}(\vec u;s)\ $$
$$  \ \ \ \ \ \ \ \ =\left( 
{\lambda\over\pi}\right)^{3\over2} 
\int e^{-\lambda \omega^2} d\omega^2 
\left( \int_{{\Bbb  R}^3} d\vec{u} \, \int_0^\infty ds\
\delta(\omega^2-(\vec p-\vec u)^2-s) {\rho}(\vec u;s) 
\right).\ \eqno(31)  $$
By comparing Eqs. (26) and (31), we see that
$(2\pi)^{3\over2} \theta(\omega) \tilde {\cal C}(\omega,
\vec p) $ and the bracket
in the r.h.s. of Eq.(31) admit the same Laplace-transform 
with respect to the variable $\omega^2 $
and are therefore equal in view of a).
Since $ \tilde {\cal C}(\omega,\vec p) $ is odd 
in the variable $ \omega $, 
Eq.(23) is therefore established and Eq.(21) immediately follows by
Fourier transformation.

\noindent
e) \sl The inversion formula (22): \rm 
by plugging Eq.(25) into Eq.(27) and the latter into Eq.(29),
we readily obtain (after an admissible interchange of integration,
the following formula being understood in the sense of
distibutions in the variables $ \vec x $ and $ s $): 
$$ {\rho}(\vec x;s) = \int_{-\infty}^
{+\infty} dx_0 \, x_0 {\cal C}(x) 
\left( {i\over2\pi} \int_{-\infty}^{+\infty} d\nu\ 
e^{{i\nu s}-{x^2\over 4i\nu}} \right) .\ \eqno (32)  $$
Since the bracket in the r.h.s. of Eq.(32) is an 
integral representation
of $ i{\partial\over\partial s} \big( \theta (s) J_0(\sqrt{sx^2})
\big) $ (see [33],
page 357, formula (11)), Eq.(32) takes the form (22) if one 
replaces $\rho $ by the distribution $ D(\vec x;m) = 
4{\pi}m\  {\rho}(\vec x;m^2)  $ (linked to $ \tilde {\rho} $ by
Eq.(24)).
\smallskip

\noindent
{}From the previous proposition, one easily deduces the following
\smallskip

\noindent
\bf Corollary:  
\smallskip

\noindent
{\it The (Fourier pair of) retarded thermal propagators
$ r(x) $ and $ \tilde r(k) $ satisfy the following
general integral representations
$$ r(x)= \int_0^{\infty} dm D(\vec x;m)\  r^{(m)}(x)  \eqno(33)  $$ 
$$ \tilde r(k) ={-1\over(2\pi)^{5\over2}} 
\int_{{\Bbb  R}^3} d\vec u \int_0^{\infty} 
ds\ {1\over{ \big( k_0^2 -(\vec k -\vec u)^2-s \big) }}\ \tilde {\rho}
(\vec u;s)\ ,\ \eqno(34)  $$
where $ k $ varies in $ T^+ $. The same representation holds for 
$ \tilde a(k) $ with $ k $ varying in $ T^- $.} 
\smallskip

\noindent
Remark: \rm

\noindent
The K\"all\'en-Lehmann representation, which is valid for a
general propagator 
in the vacuum state in the axiomatic Wightman framework 
(see [12] and references therein),  appears
as a special case of the representation (21); it is obtained
when the weight-function $ D $ is of the form 
$ D(\vec x;m)= 2m {\rho}_0{(m^2)}   $ (i.e. $ \tilde{\rho}
(\vec u;s)= (2\pi)^{1\over 2}\ {\rho}_0(s)\  \delta(\vec u) $), 
$ \rho $  being
a tempered (positive) measure with support contained in
$ \overline{{\Bbb R}^+} $. 
\vskip 1cm

\centerline{\bf 5. CONSEQUENCES OF THE KMS-CONDITION \rm} 
\parskip 5pt

\noindent The absence of the spectral condition has a drastic effect
on the structure of the $ n $-point Green functions in the space
of energy and momentum variables: as we shall explain it below,  the 
various generalized retarded functions in complex energy--momentum 
space do not admit in general a common analytic
continuation; in contrast with the case of the vacuum 
representation, \sl there do not exist 
analytic n-point Green functions in (connected)
primitive domains for the thermal
representations of quantum field theory. \rm 
It also appears as a related fact 
that \sl the (anti-) time-ordered
products and (generalized) retarded products 
have in general no mutual coincidence regions 
in real energy-momentum space,\rm\ which plagues the usual
connection between the approach in real time
and energy variables (RTF) (in terms of time-ordered products)
and the use of the Green functions in complex energy variables. 
As a matter of fact, for the case of
representations of field theory at finite 
temperature $ T={\beta^{-1}} $, the KMS-condition  
provides substitutes to the usual 
coincidence relations due to spectrum. 
These substitutes are relations which contain
the Bose-Einstein factor $ { (1-e^{-\beta\omega})}^{-1} $
and thereby imply that coincidence relations can be
only ``essentially restored, up to exponential tails'' in the
limit of very high energies.

Another consequence of the KMS-condition concerns
the justification of the approach in purely imaginary
time and energy variables.
It is a standard practice of field-theorists to use
the so-called ``Euclidean $ n $-point functions'', namely
the Wightman functions at purely imaginary times
(or Schwinger functions) and correspondingly the
$ n $-point Green functions at purely imaginary energies.
In the vacuum representation of Quantum Field Theory,
the justification of this practice is a by-product of 
the double analytic structure of the $ n $-point functions
which results from the interplay of spectral condition
and locality. In particular, it is known that the
Green functions at imaginary energies are obtained as the
Fourier transforms of the corresponding Wightman functions at
imaginary times. In the case of thermal representations of
Quantum Field Theory, a similar double analytic structure
of the $ n $-point functions also results from the interplay
of KMS-condition and locality. However, in a representation
at temperature $ T = \beta^{-1} $, the $ n $-point Wightman
functions of the complex time-variables acquire 
\it periodicity conditions with period \rm $ i\beta $ as a 
consequence of the KMS-condition. Correspondingly, the
Green functions of the complex energy-variables 
(namely the Fourier-Laplace transforms of the 
generalized retarded functions) are expected 
to be completely determined from their values on a 
certain lattice of \it discrete purely imaginary energies \rm 
of the form $ \omega ={ 2i\pi l \over\beta} $,  
with $ l $ integer, these values being
the Fourier coefficients of the corresponding ($ i\beta $-periodic)
Schwinger functions. 

These consequences of the KMS-condition have been understood to a large 
extent in the time-ordered path approach  
(see e.g. [4,7,9,23]).  
We shall exhibit them here in our general axiomatic setting 
and give complete results
for the case of the two-point function, but only indicate 
how their generalization to the $ n $-point functions
could be worked out.
\smallskip

\noindent
\bf The two-point function \rm 

\smallskip

\noindent
\sl  Substitutes to the coincidence relations in energy-momentum
space: \rm
\smallskip

\noindent 
Let $\tilde {\cal W}(p) $ be the Fourier transform of the two-point function
$ {\cal W}(x) $. The KMS-condition in the energy variable is expressed
as a special case of 
formulas (5), (7) (see Sec.1 iii')), namely:
$$  \tilde {\cal W}(p)= e^{\beta\omega}\tilde {\cal W}(-p)= 
{\tilde {\cal C}(p)\over
{1-e^{-\beta\omega}}}\ . \eqno (35)  $$
This way of writing Eq.(7) for the case of the two-point function
(with the Bose-Einstein factor at the r.h.s. of this equation)
is justified and unambiguous\it, as a relation
between positive measures, \rm in view of the additional postulate v), 
namely, the time-clustering property for ${\cal W} $.  The latter 
implies that (after the addition of a suitable constant to the field)
$ {\cal W}(x) $ tends to zero when the time-variable 
$ x_0 $ tends to infinity (at fixed $\vec x $) 
and correspondingly that the measure 
$ \tilde {\cal W}(p) $ contains no
term proportional to $ \delta(\omega) $. 
Therefore, Eq.(35) determines uniquely the splitting of the spectral
function $ \tilde{\cal C}(p)= \tilde {\cal W}(p)- \tilde
{\cal W}(-p)  $ and replaces the usual splitting
$ \tilde {\cal W}^{(vac)}(p)= \theta(\omega) \tilde{\cal
C}^{(vac)}(p)  $  which results from the spectral condition
in the vacuum representation. Eq(35) implies that $ \tilde
{\cal C}(p) $ and $ \tilde {\cal W}(p) $ ``coincide up to an 
exponential tail'' at very high energies, while
$ \tilde {\cal W} (p) $ ``vanishes up to an exponential tail''
at negative energies of very high absolute value. 

The fact that $ \tilde {\cal C}(p) $ does \it not \rm vanish
in general on any open subset of energy-momentum space 
(the examples given in Sec 2 are exceptions which will
be commented in Sec.6c below) implies that $\tilde r (p) $
and $ \tilde a (p)$  do not coincide on any open set
and therefore that the corresponding holomorphic functions 
in the tubes $ T^+ $ and $ T^- $ are \it not \rm the analytic
continuation of each other.

Let $ \tilde \tau (p) $ be the Fourier transform of the 
time-ordered function $ \tau (x) $. In view of Eqs. (12), (13)
and (35), the following relations hold: 
$$ {\tilde \tau (p)}= -i {\tilde a (p)} + {\tilde { \cal W}(p)}
={ {-i {\tilde r(p)}+i{\tilde a(p)}e^{-\beta \omega}}\over
{1- e^{-\beta\omega}}},  \eqno (36) $$ 
which show that $\tilde\tau $ and $ -i \tilde r $ only 
``coincide up to an exponential tail'' at very high energies.
\smallskip

\noindent
\sl The double analytic structure: \rm
\smallskip

\noindent The KMS-condition (see Sec. 2 iii)) implies  that there exists
a holomorphic function $ W(z_0, \vec x) $ of $ z_0 $ in the 
strip $ -\beta < Imz_0 < 0 $ whose boundary values on
the edges of this strip are respectively:
$ W(x_0, \vec x) = {\cal W}(x) $ and 
$ W(x_0-i\beta, \vec x) = {\cal W }' (x) $. Locality then implies
that the  holomorphic function $ W $ can be analytically
continued as an $ i\beta $-periodic function of $ z_0 $
in the whole complex plane minus the cuts
$ \{ z_0 = x_0 +iy_0; \left\vert x_0 \right\vert \ge 
{\vert \vec x\vert}, y_0= l\beta,   l\in {\Bbb Z} \} $. 
Moreover, the jump $ \Delta W $ of $ W $ across the 
right-hand cut $ x_0 \ge  {\vert \vec x\vert} $
is (as in the case of the vacuum representation):
$$ \Delta W(x) = \theta(x_0)\ [{\cal W}(x) - {\cal W}'(x)] 
= -i r(x).  \eqno (37) $$

If one now considers conjointly 

\noindent
i) the Fourier-Laplace transform
$$ \tilde r(k_0,\vec p) = {1\over (2\pi)^2} 
\int_{{\Bbb  R}^3}  d\vec{x} \, \int_0^\infty dx_0\ 
e^{i(k_0 x_0 -\vec p \vec x)} \ \ 
r(x_0,\vec x)\  \eqno (38) $$
of $ r $ in its holomorphy domain $ Im k_0 >0 $, and

\noindent
ii) the Fourier coefficients of the (periodic) Schwinger function
$ W(iy_0,\vec x) $,  namely 
$$  \tilde w_l(\vec p) ={1\over(2\pi)^2} 
\int_{-\beta\over 2}^{\beta\over 2} dy_0\ 
e^{{-2i\pi l\over\beta} y_0}  \int_{{\Bbb  R}^3} d\vec{x}\  
e^{-i\vec p \vec x} W(iy_0,\vec x), \eqno (39)  $$
for $ l $ non-negative integer, one can show the following basic relations:
$$ \tilde w_l(\vec p) = \tilde r({2i\pi l\over \beta},\vec p),
\ \ l= 0,\ 1,\ 2,\ldots   
\eqno (40)  $$
by means of a simple contour distortion argument. In fact,
one checks that the integral of
the product $ -(2\pi)^{-2} i e^{i(k_0 z_0 -\vec p \vec x)}
\times W(z_0,\vec x) $
over the complex cycle $ \{z_0 \in [\infty-i{\beta\over 2},
-i{\beta\over 2}]\cup [-i{\beta\over 2},i{\beta\over 2}]             
\cup [i{\beta\over 2}, \infty+i{\beta\over 2}]\} \times \{\vec x \in
{\Bbb   R}^3 \}  $ reduces to Eq (38) by shrinking the
integration cycle onto the real set $ {\Bbb  R}^+ 
\times{\Bbb  R}^3 $,
while it reduces to Eq (39) \it for the special values \rm
$ k_0 ={2i\pi l\over\beta}, $\ $ l $ integer, $ l\ge 0 $
due to the $ i\beta $-periodicity of $  W(z_0,\vec x) $.

Since $\tilde r(k_0,\vec p) $ is of moderate growth
in the complex half-plane $ Im k_0 > 0 $ (for each
$ \vec p $), it can be characterized as the
\it (unique) Carlsonian interpolation \rm 
(see e.g. [34] p.153) of the sequence of its values given by Eq.(40)
at the set of discrete imaginary energies $ {2i\pi l\over
\beta}, $\ $ l $ integer, $ l\ge 0 $; 
a corresponding result holds for the advanced
propagator $ \tilde a $ which satisfies 
in the lower half-plane equations
similar to Eq.(40) with $ l \le 0 $. 
This result together with formula (36)
make completely clear the connection between
the imaginary and real-time formalisms for the
two-point function and their equivalence in the axiomatic 
framework.
\smallskip

\noindent
\bf The $ n $-point functions \rm

\noindent The derivation of various formulas which are
substitutes to coincidence relations in energy-momentum space
relies on the following fact.  
The relations (7) implied by the KMS-condition
can be extended to a more general set of similar
formulas, namely
$$ (1- e^{-\beta\omega_I})\  
\langle \Omega_\beta,\tilde \Pi(I)\tilde \Pi(J) \Omega_\beta \rangle
=\langle \Omega_\beta,[\tilde\Pi(I), \tilde\Pi(J)] \Omega_\beta \rangle, 
\eqno (41) $$
where $ \tilde\Pi(I) $ denotes the Fourier transform 
of any (product of) generalized
retarded or {(anti)} time-ordered products of field operators 
depending on the energy-momentum variables  $ p_i=
(\omega_i,\vec p_i)$, $i\in I $. 
Since all the distributions involved are tempered, it follows
from Eq.(41) that all the expressions of the form
$ \langle \Omega_\beta, \tilde\Pi(I) \tilde\Pi(J) \Omega_\beta \rangle $
decrease exponentially as $ e^{-\beta
\vert\omega_I\vert} $ for $\omega_I $ tending to $ -\infty $. 

By taking the latter property into account for all the terms
of the sum at the r.h.s. of Eq.(20) (after applying a
Fourier transformation to both sides of this equality),  
one easily checks the following statement:

\noindent
\sl Consider for each cell $ \alpha $ 
the region $ \hat\gamma_\alpha $ of (real) energy--momentum 
space which is defined by the condition
$ \omega = (\omega_1, \ldots ,\omega_n) \in \gamma_\alpha $;
then, in this region $ \hat\gamma_\alpha $, 
the corresponding distribution
$ \tilde r_\alpha(p) $ ``essentially coincides'' with
the Fourier transform $ \tilde {\tau}(p) $ of the time-ordered
expectation value 
$ \langle \Omega_\beta, T(\{1,2,\ldots,n \}) \Omega_\beta
\rangle $ up to terms
which are exponentially decreasing with respect to the
various energy variables inside $ \hat \gamma_\alpha $. \rm

Formula (41) also applies to the r.h.s. of the following relations
which result from Eq.(19) by a Fourier transformation:
$$ \tilde r_{\alpha_1} -\tilde r_{\alpha_2} =
\langle \Omega_\beta,[\tilde R_{\alpha^{(I)}},\tilde R_{\alpha^{(J)}}]
\Omega_\beta \rangle\ .\ \eqno(42) $$
However, in contrast with what happens for the vacuum representation
(in view of the spectral condition), formula (41) does not imply
any support property for the commutator function at the r.h.s. of
Eq.(42). Therefore, \sl the holomorphic functions $\tilde r_{\alpha_1}(k) $
and $ \tilde r_{\alpha_2}(k) $ whose boundary values on the reals
are linked by Eq.(42) do not admit mutual analytic continuation
via the edge-of-the-wedge property \rm as it is the case 
for the Green functions of the vacuum representation.

The analytic structure of the $ n $-point functions in the complex
time-variables results from the KMS-conditions (see Eqs. (3), (4))
which one writes for \it all the permuted Wightman distributions 
\rm $ {\cal W}_n^{(\beta)}(I,J) $, $ (I,J) $ denoting any pair of
ordered sets forming an ordered partition of $ \{1,2,\ldots,n\} $.
As in the case of the vacuum representation, 
there exists (for each $ n $) a distribution
$ W_n^{(\beta)}((z_{1,0},\vec x_1),\ldots, (z_{n,0},\vec x_n)) $ which is
holomorphic in a union of permuted tubes in the space of 
complex time-variables $ z_{i,0}, \ \ 1\le i\le n $, together
with regions of mutual crossover provided by locality.

However, what is different and characteristic of the
representations at temperature $ \beta^{-1} $ is the fact
that the tubes have \it bounded \rm polyhedral bases which 
form a multiperiodic paving of the space of purely imaginary
time-variables $ y_0=(y_{1,0},\ldots,y_{n,0}) $.
In particular, the holomorphy domain of the $ W_n^{(\beta)} $
contains all the Euclidean configurations $ z_j=(iy_{j,0},\vec x_j),
\  1\le j\le n, $ except those which satisfy conditions
of the form $ z_j =z_k + il\beta e_0,\  l\in {\Bbb Z}, $
for all possible pairs $ (j,k) $.
One therefore also expects a generalization to take place
concerning \it the double analytic structure in the time and 
energy variables \rm which we have exhibited above for the 
two-point function. More precisely, one expects the derivation of 
formulas of the form (40) to be feasible in the general case;
such formulas would relate the values
of the generalized retarded $ n $-point functions 
$ \tilde r_\alpha $ on appropriate lattices
of purely imaginary energies to the Fourier coefficients
of the Schwinger $ n $-point functions which should be computed 
(for each $ r_\alpha $) on a corresponding periodicity pattern
of $ y_0 $-space. This
is an algebraic problem
which can be solved rather easily for $ n $=3 and is expected to be
solvable for general $ n $ with some amount of work [35].  
\vskip 1cm

\centerline{\bf 6. COMPLEMENTS AND DISCUSSION \rm}
\parskip 5pt 

\noindent
\bf a) Integral representation of the two-point correlation function
\rm

\noindent It follows from Eqs. (14) and (35) that the Fourier transform
of the two-point correlation function of the free field
at temperature $\beta^{-1} $ is
$$ \tilde {\cal W}_{\beta}^{(m)}(\omega,\vec p) =
{1\over 2\pi}
{\epsilon(\omega)\delta(\omega^2 - {\vec p}\ ^2 - m^2)\over
1- e^{-\beta\omega}} .    \eqno (43)  $$

By multiplying both sides of Eq.(23) by $(1- e^{-\beta\omega})^{-1} $
and taking Eq.(35) into account, one then obtains the following
general representation for the Fourier
transform of the two-point correlation function
of any local scalar field at temperature $ \beta^{-1} $:
$$  \tilde {\cal W}(\omega,\vec p) =
{1\over(2\pi)^{3\over2}} \int_{{\Bbb R}^3} d\vec u \int_0^{\infty} 
ds\  \tilde \rho(\vec u,s)\  \tilde {\cal W}_{\beta}^{(\sqrt s)} 
(\omega,\vec p -\vec u).   \eqno (44) $$  

By coming back to the space-time variables, we obtain correspondingly 
the following integral representation for the correlation function
itself, which is comparable to formula (21): 
$$ {\cal W}(x) = \int_0^{\infty} dm\  D(\vec x;m)\   {\cal W}
_{\beta}^{(m)}(x).    \eqno (45)  $$ 

\noindent
\bf Remark: \rm

\noindent The consideration of ``generalized free thermal fields''
completely defined by specifying their two-point function 
(and imposing the prescription that all the truncated $ n $-point
functions, with $ n > 2 $, should vanish) is now possible.
We can say that all the information which is necessary
to construct such a field satisfying all the general
principles of Thermal Quantum Field Theory defined above
(see Sec.2) is encoded in the ``weight-function'' $ D(\vec x;m) $
(or $ \tilde \rho(\vec u,s) $) of the integral representation
(45) (or (44)) of its two-point function; the only restriction
which the tempered distribution $ \tilde \rho(\vec u,s) $ has
to comply with is the following one: the convolution expressed (formally)
by the double integral at the r.h.s. of Eq. (23)  must give a 
positive measure. This is clearly the case if 
$\tilde \rho ( \vec{u}, s )$ happens to be positive. 

\noindent
\bf b) Consequences of the relativistic KMS-condition \rm 

\noindent 
The implications of the relativistic KMS-condition (presented above 
in Sec.2, conditions iv), iv')) concern the existence of an 
\it analytic structure of the $ n $-point functions in the 
complex space-time vector variables \rm and the exploitation of
corresponding \it exponential decay properties 
in the space of (real) energy-momentum variables. \rm 
The latter are produced  not
only at very large negative energies
but also at very large momenta; they in fact apply to
energy-momentum regions and to distributions  for which  
the relativistic spectral condition would imply 
vanishing properties (in the vacuum representation of field
theory).

For the case of the two-point function, a detailed study  
will be given in [32] with the following type of results. 
Even under the weakest form of relativistic KMS-condition (see [17]),
the function $ W(z_0,\vec x) $ introduced in Sec.5 admits
an analytic continuation $ W(z) $ as a holomorphic function
of the complex four-vector $ z=(z_0, \vec z) $ in a domain
which contains the $i\beta $-periodic ``flat cut-domain'' of $ W $ 
in $ (z_0, \vec x) $-space. This shows (as a by-product) the
regular (i.e. $ {\cal C}^{\infty} $) character of the
distribution $ {\cal W}(x_0,\vec x) $ with respect to the 
spatial coordinates $ \vec x $. Moreover, the distribution 
$ D(\vec x,m) $ appearing as a ``weight-function'' 
in the representations (21) and (45) is also shown to have a
${\cal C}^{\infty} $-dependence in $\vec x $.
Under the strongest form of relativistic KMS-condition
(namely, the one presented as condition iv) in Sec.2),
$ W(z) $ is holomorphic in the union of the tube $ T_{\beta} $
together with all those which are obtained from the latter
by the translations $ il{\beta}e_0,\  l \in {\Bbb Z} $.  
All these tubes are connected together by 
complex neighbourhoods of the  
regions $ \{z;\  y_0 = l{\beta},\  x_0^2 - {\vec x}^2 <0, 
\ l \in {\Bbb Z} \} $ given by locality. The weight-function
$ D $ is also proved to admit an analytic continuation
$ D(\vec z,m) $ in a corresponding tube with imaginary basis
(containing the origin) 
in the space of complex coordinates $ \vec z $. 
In the energy-momentum variables $ (\omega,\vec p) $,
the product of the spectral function $ \tilde {\cal C}(p) $ 
by the function $ e^{{\beta\over 2}[(1+{\vec p}^2)^{1\over 2}
-(1+ \omega^2)^{1\over 2}]} $ must be a tempered distribution 
and the ``exponential tail'' of $ \tilde \tau +i\tilde r $
extends to the whole complement of the forward cone
in $ {\Bbb R}^4 $; correspondingly, the ``weight-function''
$ \tilde \rho(\vec u,s) $ of the representation (22)
is exponentially decreasing with respect to $ \vert \vec u \vert $. 
Similar properties could be derived for the $ n $-point functions.

\noindent
\bf c) Feynman-type operations on two-point functions and
discussion of examples \rm

\noindent The double analytic structure of thermal two-point functions
which we have described (see Sec.5) can be shown to be
preserved under the two basic Feynman-type operations, 
namely

\noindent
\sl i)``$ N $-line Wick-contraction''\rm\  (represented by the 
diagram with two vertices $ z_1 $ and $ z_2 $ connected by $ N $
lines):

\noindent With such a diagram $ \Gamma $ is 
associated the product $ W_{[\Gamma]}(z_1-z_2) $ of $ N $
two-point functions taken at the same complex point $ z_1- z_2 $
(each factor $ W_{(\lambda)}(z_1-z_2) $ being 
associated with a line $ \lambda $). 
It is clear that such a product of holomorphic functions still
satisfies the same properties ($  i\beta $-periodicity, 
analyticity domain expressing the relativistic KMS-condition,
boundary values on the reals in the sense of distributions, locality) as each 
individual factor. These properties characterize $ W_{[\Gamma]} $
as being a thermal two-point function whose associated propagator
$ \tilde r_{[\Gamma]}(k) $ or $ \tilde a_{[\Gamma]}(k) $ 
can be computed on the discrete sequence
of energies $ \omega_l = i{2\pi l\over \beta},\ l\in {\Bbb Z},  $ 
(see Eq.(40)) 
as a (discrete) convolution of the $ N $ propagators $ \tilde
r_{(\lambda)} $ or $ \tilde a_{(\lambda)} $ 
taken on the same sequence of imaginary energies. 
Moreover, the Fourier transform of the correlation function  
$ \tilde {\cal W}_{[\Gamma]}(p) $ is also the convolution
product on the (real) energy-momentum space of the 
Fourier transforms of the $ N $ correlation functions
$ \tilde {\cal W}_{(\lambda)} $.

\noindent
\sl ii) ``Vertex convolution''\rm: 

\noindent The vertex convolution of the thermal two-point
functions $ W_{(1)} $ and $ W_{(2)} $ is defined as the following 
convolution product on Euclidean space-time:\ \ \  
$ W(z_1-z_2) = \left( W_{(1)}* W_{(2)}\right)(z_1-z_2) =
\int_{-i{\beta\over 2}}^{i{\beta\over 2}} dz_0 \int_{{\Bbb R}^3}d\vec z
\ W_{(1)}(z_1-z)\ W_{(2)}(z-z_2). $ The resulting holomorphic function  
$ W $ still satisfies all the structural properties of thermal two-point
functions and the associated propagator $ \tilde r(k) $ in  complex 
energy-momentum space is the ordinary product of the 
propagators $ \tilde r_1(k) $ and $ \tilde r_2(k) $  
associated with $ W_1 $ and $ W_2 $.

In particular, if one starts from the free thermal two-point
functions  (see Sec.3, Example 1, and Sec.6a)), one can construct
as an application of the operation i)  appropriate perturbative 
two-point functions associated with ``self-energy bubble diagrams'' 
of the thermal interacting field. Then, the operation ii) defines  
in a rigorous way the two-point function obtained by ``resummation of the 
corresponding iterated self-energy diagrams''. If the self-energy
contribution is denoted (in complex energy-momentum space) by 
$ \Sigma (k) $, the associated ``complete'' propagator is given,
as usual, in the complex domains ($ T^+ $ and $ T^- $) by the 
formula $ \Delta (k) = {1 \over k_0^2 - {\vec k}^2 - m^2 - \Sigma (k)} $.

All these structural properties could of course be derived
similarly for non-scalar fields such as those of QCD. One can thus 
understand in this general framework such computations of gluon propagators
as those given by Weldon and Pisarski [5,6] (see Sec.3, Example 2). 
In the latter, however,
a high-temperature approximation has been taken which makes the
expressions (16) and (17) 
of the propagators $ \Delta_l $ and $ \Delta_t $ somewhat peculiar, 
as far as the support and decrease properties of the associated
spectral functions are concerned.

We first notice that, apart from those of the free field  
and of the two-lined bubble diagram, the spectral functions of 
all the perturbative two-point functions generated by the
previous operations i) and ii) have \it no support restrictions; \rm 
however, they all enjoy a property of exponential decrease
at large momenta which expresses the fact that the relativistic
KMS-condition is satisfied.  These two properties are violated
by the spectral functions of the WP-propagators, since:

\noindent
a) the latter only satisfy a condition of power decrease at large momenta: 
but this ``hard thermal loop approximation'' is generally used only 
in the low momentum region;

\noindent
b) in the usual relativistic spectral region $ \vert \omega \vert
> \vert \vec p \vert $, their support  is restricted to a pair of  
hypersurfaces of the form $ \omega = \pm \omega_{t,l}(\vec p) $ 
which correspond to sharp terms containing the factors 
$ \delta(\omega\mp \omega_{t,l}(\vec p)) $;  the existence of
such real dispersion laws, interpreted as ``plasmon modes'', 
calls for further comments which open our last topic. 

We conclude these remarks by noting that a general perturbative
approach to the construction of thermal correlation functions,
based on the systematic exploitation of locality and KMS-condition,
has been proposed by Steinmann [36].

\noindent
\bf d) Representations of particles \rm

\noindent The concept of \ \it ``dispersion law'' \rm
$ \omega = \omega_{part}(\vec p) $ 
\ for representing a ``particle'' {(or ``mode'')} in Thermal Quantum
Field Theory may seem the most natural one to be inherited from the 
familiar formalism of quantum field theory in the vacuum
representation, having in mind that the hypersurface defined by
this law in the space of the energy-momentum 
variables might not be (in 
general) a relativistic hyperboloid shell.
Such a hypersurface should appear as 
the singular set of a \it real pole \rm in the thermal 
propagators of the theory, or equivalently 
as the support of a \it $ \delta $-term \rm 
in the corresponding spectral functions.  
However, this attractive picture turns out to be \it false, except
precisely in the trivial case of a relativistic free particle
moving across the thermal bath without any interaction: \rm so is the 
content of the Narnhofer-Requardt-Thirring Theorem [37] 
which has been proved by these authors in the general 
framework of ``Local Quantum Physics'' [22].
This frustrating result holds true, even if one allows the support of 
the $ \delta $-term to be imbedded in a region where 
the spectral function has a non-zero continuous background. 
As a matter of fact, this is not so surprising since
a $ \delta $-term in $ \tilde {\cal C}(p) $ would imply 
that the correlation function
$ {\cal W}(t,\vec x) $ has the ``normal'' slow decay
property (as $ \vert t \vert^{-{3\over2}} $) along any world-line 
$ {\vec x} = {\vec v}t $, and therefore that the particle is not
submitted to
dissipative effects due to the interactions with the thermal bath.
For example, the WP-gluon-propagators considered above represent an
approximation which does not provide a realistic description
in terms of particles in a thermal equilibrium state. 

A reasonable way out of this deadlock, which has been proposed in
particular by Landsman [38], consists in adopting the same viewpoint as
for the representation of unstable particles in the vacuum state of
field theory. This amounts to assume that the retarded propagator
(or a distinguished part of it)
admits an analytic continuation from the upper $ k_0 $-plane
(or the tube $ T^+ $) into the  
lower $ k_0 $-plane, across (part of) the reals, and that a
\it complex pole \rm $ k_0 = \omega_{part}(\vec p) -i \gamma_{part}
(\vec p) $ is present in this \it second-sheet \rm domain.
Of course, this \it complex dispersion law \rm should  
present some characteristics which would distinguish the 
type of decay (due to statistical dissipative effects) 
of a particle in a thermal
bath from the one (due to intrinsic unstability) of a resonance. 
In particular, one would expect the law to be 
such that the dissipative effects are barely 
felt by the particle \it at rest, \rm  
namely that $ {\cal W}(t,{ \vec v}t) $ keeps the $ \vert t
\vert^{-{3\over 2}} $ behaviour for $ \vec v = 0 $, while
being exponentially decreasing in $ t $ for $ \vec v \ne 0 $.
This is certainly not the case for the simplest ways of choosing
the ``width'' $ \gamma_{part}( \vec p) $ (i.e. constant or proportional to
$ \omega_{part}(\vec p) $). 

At this point, we would like to advocate 
(as in [18,19])
that our general integral 
representations (23) and (45) of 
the thermal spectral functions and correlation
functions provide a somewhat more natural prescription for 
representing particles: let us assume that,
as in the K\"all\'en-Lehmann formula for the spectral function in the vacuum
state, a particle is associated with a \it $ \delta $-term in the
weight-function $ \tilde \rho(\vec u, s) $ of  
the representation (21),\ \rm \ namely, with a term of the form
$ \tilde \rho_{part}(\vec u)\  \delta(s-m_0^2) $. 
The representation (45) then contains the distinguished term
$ D_{part}(\vec x)\ {\cal W}_{\beta}^{(m_0)}(x) $, where
$ D_{part}(\vec x) $ is (up to a factor) the Fourier transform of
$ \tilde \rho_{part}(\vec u) $. It is now clear that, if we
interpret $ D_{part}(\vec x) $ as a \it ``dissipative (or damping) 
factor'' \rm
which should decrease, for instance, exponentially at large
$ \vec x $, this distinguished term is a good candidate for
representing a particle behaviour. (Note that it would satisfy the
exponential decrease property along all world-lines 
$ \vec x = {\vec v}t $, but would behave as 
$ \vert t \vert^{-{3\over2}} $
at rest, like the free correlation 
function $ {\cal W}_{\beta}^{(m_0)}$.) 
Since in this case, the function $ \tilde \rho_{part} $  
would be holomorphic in a tube domain of the complexified variables 
$ \vec u $, one can prove that the corresponding propagator represented by
formula (34) would then have analytic continuation properties in
the lower half-plane of the variable $ k_0 $. 
This indicates (and a further general study confirms it 
as well as the special example presented in [18,19]) that this
\it $ \delta $-representation of particles, \rm \  
far from being contradictory with the 
previous one based on ``second sheet complex poles''
of the propagators, \it selects inside the latter class \rm  
candidates which might be the most appropriate representatives 
of the notion of particle in 
Relativistic Thermal Quantum Field Theory.  

Finally, one can also show that the integral 
representations (21) and (23) are a useful tool for investigating
the manifestations of ``thermal Goldstone particles'' 
in the case of ``spontaneous symmetry breaking'';
this will be the object of a further work [39]. 

In conclusion, we have presented a general framework ``a la Wightman''
for the concepts of Thermal Quantum Field Theory and 
a set of (preliminary) results, 
mainly expressed in terms of analytic structural properties of
the $ n $-point functions of the fields. 
These results are already rather complete for the case $ n=2 $. 
They are useful for clarifying some structural aspects of
perturbative results, obtained in the previous years 
by various authors. Moreover, the general integral representation
of thermal two-point functions 
that we have obtained is interesting under several respects:
providing a complete knowledge 
of ``generalized free thermal fields'', 
it also suggests a rather promising 
approach to the concept of particle, integrating the previously
known ideas and consistent with the general principles
of relativistic thermal field theory.
Many open problems remain, as far as the properties and the use 
of $ n $-point functions are concerned (for $ n> 2 $); among the 
important ones, let us mention in particular the problem of the 
possible definition of several-particle states as a manifestation of
the field interactions in a spirit comparable to 
the collision theory in the vacuum state. 	
\vskip 0.8cm

\noindent
\centerline {\bf ACKNOWLEDGEMENTS \rm} 
\parskip 5pt

We wish to thank P.A.\ Henning for useful discussions. 
Financial support of this research project by 
the Franco-German science
cooperation PROCOPE is gratefully acknowledged. 
\vskip 0.8cm

\noindent
\centerline {\bf REFERENCES \rm} 
\parskip 5pt

\noindent
[1] Proceedings of the International Symposium on Statistical 
Mechanics of Quarks and Hadrons, Bielefeld 1980;
ed.\ H.\ Satz (North Holland, Amsterdam 1981)  

\noindent
[2] Proceedings of the 3rd Workshop on Thermal Field Theories and 
Their Applications, Banff 1993; 
eds.\ F.C.\ Khanna, R.\ Kobes, G.\ Kunstatter and H.\ Umezawa (World  
Scientific, Singapore 1994) 

\noindent
[3] Proceedings of the NATO Advanced Research Workshop on
Hot Hadronic Matter, Divonne 1994; 
eds.\ H.\ Gutbrod, J.\ Letessier and J.\ Rafelski (Plenum, New York 1995)

\noindent
[4] N.P.\ Landsman and Ch.G.\ van Weert, \it Phys.\ Reports \bf 145 \rm
(1987) 141

\noindent
[5] H.A.\ Weldon, \it Phys.\ Rev.\ \bf D26 \rm (1982) 1394, and
\it Phys.\ Rev.\ \bf D47 \rm (1989) 2410

\noindent
[6] R.\ Pisarski, \it Physica \bf A158 \rm (1989) 146

\noindent
[7] P.A.\ Henning, \it Phys.\ Reports \bf 253 \rm (1995) 235 

\noindent
[8] M.\ Lutz and T.\ Kunihiro, \it Z. Phys. \bf C49 \rm (1991) 123
and R.L. Kobes, \it Z. Phys. \bf C53 \rm (19920 537

\noindent
[9] T.S.\ Evans, \it Phys.\ Rev.\ \bf D47 \rm (1993) 4196 
{\it and} T.S.\ Evans and A.C.\ Pearson, \it Phys.\ Rev.\ \bf D52 \rm (1995)
4652

\noindent
[10] R.F.\ Streater and A.S.\ Wightman, \it PCT, Spin and Statistics and
All That, \rm Benjamin, New York (1964) 

\noindent
[11] R.\ Jost, \it The General Theory of Quantized Fields, \rm American
Math.\ Society, Providence (1965)

\noindent
[12] N.N.\ Bogolubov, A.A.\ Logunov, A.I.\ Oksak and I.T.\ Todorov,
\it General Principles of Quantum Field Theory, \rm Nauka
Publishers, Moscow (1987) and Kluwer Academic Publishers, Dordrecht
(1990)

\noindent
[13] D.\ Iagolnitzer, \it Scattering in Quantum Field Theories, \rm 
Princeton University Press, Princeton (1993)

\noindent
[14] R.\ Haag, N.M.\ Hugenholtz and M.\ Winnink, \it Commun.\ Math.\ Phys.\
\bf 5 \rm (1967) 215

\noindent
[15] R.\ Haag, D.\ Kastler and E.B.\ Trych-Pohlmeyer, \it Commun.\ Math.\
 Phys.\ \bf 38 \rm (1974) 173

\noindent
[16] W.\ Pusz and S.L.\ Woronowicz, \it Commun.\ Math.\ Phys.\ \bf 58 \rm 
(1978) 273

\noindent
[17] J.\ Bros and D.\ Buchholz, \it Nucl.\ Phys.\ \bf B429 \rm (1994) 291

\noindent
[18] J.\ Bros and D.\ Buchholz, \it Z.\ Phys.\ \bf C55 \rm (1992) 509

\noindent
[19] J.\ Bros and D.\ Buchholz, \it Relativistic KMS-Condition and
K\"all\'en-Lehmann Type Representations of Thermal Propagators, 
\rm in:   
Proceedings of the 4th Workshop on Thermal Field Theories and 
their Applications, Dalian 1995 (to be published) and  hep-th/9511022 

\noindent
[20] H.J.\ Borchers, \it Nuovo Cimento 
\bf 24 \rm (1962) 214 

\noindent
[21] A.\ Uhlmann, \it Wiss.\ Zeits.\ Karl Marx Univ.\ \bf 11 \rm (1962) 213 

\noindent
[22] R.\ Haag, \it Local Quantum Physics, \rm Springer, Berlin (1992) 

\noindent
[23] T.S.\ Evans, \it N-point functions in imaginary and real time 
finite temperature formalisms 
\rm in: \it Proceedings of the 2nd Workshop on Thermal Field Theories
and their Applications, \rm Tsukuba 1990; ed.\ H.\ Ezawa et al.\  
(North Holland, Amsterdam 1991)  and  
\it Nucl.\ Phys.\ \bf B374 \rm (1992) 340 

\noindent 
[24] O.\ Steinmann, \it Helv.\ Phys.\ Acta \bf 33 \rm (1960) 257 

\noindent
[25] H.\ Araki and N.\ Burgoyne, \it Nuovo Cimento \bf 18 \rm (1960) 342 
and H.\ Araki, \it J.\ Math.\ Phys. \bf 2 \rm (1961) 163

\noindent
[26] D.\ Ruelle, \it Nuovo Cimento \bf 19 \rm (1961) 356

\noindent
[27] J.\ Bros, Thesis, Paris (1970) 

\noindent
[28] H.\ Epstein, V.\ Glaser and R.\ Stora,  
\it General properties of the n-point functions in local quantum field
theory, \rm in: \it Structural analysis of collision amplitudes,
\rm eds.\ R.\ Balian and D.\ Iagolnitzer , North Holland, Amsterdam (1976) 
pp.\ 7-94

\noindent
[29] A.S.\ Wightman, \it Analytic Functions of several complex variables, 
\rm in: \it Dispersion Relations and Elementary Particles, \rm Wiley, New
York (1960) pp. 159-221

\noindent
[30] J.\ Bros, P.A.\ Henning, E.\ Poliatchenko and T.\ Schilling,
\it Approximate spectral functions in thermal field theory, 
\rm to appear

\noindent
[31] J.-L.\ Gervais and F.J.\ Yndurain, \it General Representation of 
Causal Distributions, \rm unpublished 

\noindent
[32] J.\ Bros and D.\ Buchholz, \it Fields at finite temperature:
A general study of the two-point functions, \rm in preparation

\noindent
[33] H.\ Bateman, \it Tables of Integral Transforms Vol.I, \rm  
McGraw-Hill, New York (1954)

\noindent
[34] R.P.\ Boas, \it Entire Functions, 
\rm Academic Press, New York (1954)

\noindent
[35] J.\ Bros, work in progress 

\noindent
[36] O.\ Steinmann, \it Commun.\ Math.\ Phys.\ \bf 170 \rm (1995) 405
and \it Ann.\ Inst.\ H.\ Poincar\'e \bf 63 \rm (1995) 399 

\noindent
[37] H.\ Narnhofer, M.\ Requardt and W.\ Thirring, \it Commun.\ Math.\ Phys.\
\bf 92 \rm (1983) 247

\noindent
[38] N.P.\ Landsman, \it Ann.\ Phys.\ \bf 186 \rm (1988) 141

\noindent
[39] J.\ Bros and D.\ Buchholz, \it The Unmasking of Thermal
Goldstone Bosons, \rm to appear

\end